\documentclass[prl,aps,notitlepage,twocolumn,longbibliography,superscriptaddress,preprintnumbers]{revtex4-1}
\usepackage[dvipsnames, svgnames]{xcolor}
\usepackage{amsfonts,amssymb,amsmath,mathrsfs,graphicx,bm,orcidlink,url,here,mathtools,chngcntr,ulem}
\usepackage{hyperref}
\usepackage[whole]{bxcjkjatype}
\newcommand{\secl}[1]{\noindent \textbf{#1—}}
\newcommand{\MM}{\mathrm{MM}}

\begin{document}

\title{Exact Cancellation of Horizon Modes in Kerr Ringdown}

\preprint{RESCEU-23/26}

\author{Kei-ichiro Kubota\orcidlink{0000-0002-1576-4332}}
    \affiliation{Institute for Cosmic Ray Research, The University of Tokyo, 5-1-5 Kashiwanoha, Kashiwa, Chiba 277-8582, Japan}
\author{Yusuke Manita\orcidlink{0000-0002-2673-8404}}
    \affiliation{Perimeter Institute for Theoretical Physics, Waterloo, Ontario, N2L 2Y5, Canada}
    \affiliation{Department of Physics, Tokyo Metropolitan University, 1-1 Minami-Osawa, Hachioji, Tokyo 192-0397, Japan}
\author{Hayato Motohashi\orcidlink{0000-0002-4330-7024}}
    \affiliation{Department of Physics, Tokyo Metropolitan University, 1-1 Minami-Osawa, Hachioji, Tokyo 192-0397, Japan}
\author{Daiki Watarai\orcidlink{0009-0002-7569-5823}}
    \affiliation{Research Center for the Early Universe (RESCEU), Graduate School of Science, The University of Tokyo, Tokyo 113-0033, Japan}
\date{\today}

\begin{abstract}
Ringdown gravitational waves probe black hole spacetimes via quasinormal modes (QNMs). 
Recent studies have reported oscillatory features in QNM-filtered waveforms, and have led to conflicting interpretations as to whether they represent a near-horizon ``direct wave.''
Reexamining a particle plunging into a Kerr black hole, we show that the saddle-point approximation underlying the proposed interpretation fails near the horizon.
Our exact analysis reveals that every source-induced horizon-mode pole is canceled by an infinite tower of Matsubara zeros, independently of orbital and black-hole parameters.
The first four stages of this sequential screening are also confirmed by time-domain numerical integration.
Thus, within the near-horizon perturbative framework, source-induced horizon-mode signals do not survive at late times.
\end{abstract}

\maketitle  

\secl{\label{sec:introduction}Introduction}Probing gravity in the strong-field regime is one of the ultimate frontiers in modern physics.
Gravitational waves emitted during black-hole ringdown offer a valuable probe of spacetime geometry in this regime.
Within linear perturbation theory, the ringdown waveform is conventionally described as a superposition of damped sinusoids, known as quasinormal modes (QNMs), together with a late-time power-law tail~\cite{Berti:2025hly}.

Recent studies have focused on oscillatory features in waveforms obtained after applying the QNM filter~\cite{Ma:2022wpv}, and their origin and interpretation remain under active debate~\cite{Oshita:2025qmn,Lu:2025vol,Kankani:2026byb,Chung:2026eph,Kuntz:2026xep,Dyer:2026yex,Kankani:2026kst,Ma:2026hcb,Su:2026gmp,Sun:2026mto,Cheung:2026gfd}.
Within black-hole perturbation theory, Oshita \textit{et al.}~\cite{Oshita:2025qmn} used the saddle-point approximation (SPA) to propose a ``direct wave'' model, in which a contribution sourced by the plunge is assumed to reach infinity more directly than the light-ring QNM response.
In this model, the QNM-filtered waveform was argued to evolve toward a late-time horizon-mode contribution~\cite{Mino:2008at,Zimmerman:2011dx} whose frequency is determined by the remnant horizon~\footnote{Related source-dependent contributions considered for Schwarzschild black hole, termed ``redshift modes''~\cite{DeAmicis:2025xuh,Rosato:2026moe,DeAmicis:2026wqd}, share the same characteristic frequency and have been shown to vanish identically in Schwarzschild spacetime~\cite{Kuntz:2026xep}.
We focus here exclusively on horizon modes, whose cancellation mechanism is distinct from the redshift-mode result of Ref.~\cite{Kuntz:2026xep}.}.
They reported features in plunging-particle waveforms~\cite{Kojima:1984cj} that they interpreted as signatures of this model, and extended this interpretation to SXS numerical-relativity waveforms of comparable-mass black-hole mergers~\cite{Boyle:2019kee}.
If correct, this interpretation would have substantial observational implications.
It has motivated searches for a proposed ``direct wave'' component in GW250114~\cite{LIGOScientific:2025rid} and proposals to use such a component to probe near-horizon physics~\cite{Lu:2025vol,Chung:2026eph}.

However, the recent literature has not converged on a common physical interpretation. 
Although the oscillatory features in the filtered waveform have been interpreted as a ``direct wave,'' their frequencies agree with the direct-wave prediction only for particular spins, with noticeable discrepancies for others~\cite{Ma:2026hcb,Kankani:2026byb,Kankani:2026kst,Sun:2026mto}.
These discrepancies may reflect intrinsic limitations of the direct-wave model, in particular its reliance on the SPA, whose applicability has been explicitly questioned in a Schwarzschild study~\cite{Kuntz:2026xep}.
However, the behavior of the proposed late-time contribution in Kerr without this approximation has not yet been established.

In this Letter, we reexamine the late-time horizon-mode limit assumed in the proposed ``direct wave'' construction for a point particle plunging into a Kerr black hole.
From a viewpoint different from that of Ref.~\cite{Kuntz:2026xep}, we find that the SPA used in this construction fails in the near-horizon limit.
While under the SPA the candidate contribution asymptotes to the horizon-mode frequency~\cite{Oshita:2025qmn,Sun:2026mto}, a simple exact evaluation of the source integral without the SPA instead reveals that an infinite tower of Matsubara zeros exactly cancels the would-be horizon-mode contributions, regardless of the detailed orbital profile and of the values of the black-hole and trajectory parameters.
This result is consistent with the recent finding that QNM filtering redistributes QNM power in time and that the oscillatory features near the peak of the filtered waveform are dominated by radiation sourced at or outside the light ring~\cite{Cheung:2026gfd}.
Our result rules out the proposed ``direct wave'' interpretation of the oscillations in the filtered waveform as a surviving late-time horizon-mode signal.
Throughout this Letter, we focus on the dominant multipole with $(\ell,m)=(2,2)$ and adopt units with $c=G=1$.

\secl{\label{sec:horizonmode}Horizon-mode construction and screening}We first review the source-induced horizon mode~\cite{Mino:2008at} and its screening~\cite{Zimmerman:2011dx}.
Within the Teukolsky formalism, Mino and Brink identified a source-induced pole in the frequency-domain gravitational-wave amplitude for a particle plunging into a Kerr black hole~\cite{Mino:2008at}.
Unlike a QNM pole, this pole originates from the source integral rather than from the Green's function, and the resulting candidate contribution is referred to as a horizon mode.
The horizon-mode frequency $\omega = m\Omega_\mathrm{H} - 2\pi i T_\mathrm{H}$ is literally determined by the quantities at the event horizon $r_+=M+\sqrt{M^2-a^2}$ of a Kerr black hole with mass $M$ and spin parameter $a$.
Here, $\Omega_\mathrm{H}=a/(2Mr_+)$ is the angular velocity at the horizon and $T_\mathrm{H}=\kappa/(4\pi r_+)$ is the Hawking temperature, where $\kappa=\sqrt{1-(a/M)^2}$~\footnote{Note that this definition of $\kappa$ differs from that of the conventional surface gravity.}.
Therefore, if the horizon mode is observable, it would offer an unprecedented opportunity to probe near-horizon physics that remains inaccessible via QNMs, which characteristically reflect the properties of the light ring.

At the same time, the horizon-mode frequency coincides with the fundamental mode ($j=1$) of the Matsubara frequencies
\begin{align}
    \omega^\mathrm{MM}_j = m\Omega_\mathrm{H} - 2\pi i T_\mathrm{H}j, \qquad j=1,2,\dots,
    \label{eq:MatsubaraModeFrequency}
\end{align}
whose nontrivial roles in black-hole physics have recently been highlighted~\cite{Kuntz:2025gdq,Arnaudo:2025uos,Motohashi:2026mbn,Kubota:2026hdv}.
As we show below, they play an additional role by enforcing the exact cancellation of all source-induced horizon-mode contributions.

Let us briefly review how the horizon mode arises from the source term.
We consider a gravitational wave emitted by a particle plunging into a Kerr black hole along a timelike geodesic. 
In the near-horizon region $\epsilon = (r-r_+)/r_+ \ll 1$, the geodesic equations in the Boyer-Lindquist coordinates yield the trajectory of the plunging particle as
\begin{align}
    \epsilon(t) &= \frac{r(t)-r_+}{r_+}=e^{-\kappa(t-t_0)/r_+} = e^{2\mathrm{Im}(\omega^\MM_{1})(t-t_0)},\label{eq:horizongeodesicr} \\
    \theta(t) &= \theta_0 +\mathcal{O}(\epsilon)\label{eq:horizongeodesictheta}, \\
    \phi(t)&=\Omega_\mathrm{H}(t-t_0) + \phi_0 + \mathcal{O}(\epsilon) \label{eq:horizongeodesicphi}, \\
    r_*(t) &= -t + \mathcal{O}(\epsilon), \label{eq:horizongeodesicrstar}
\end{align}
where $r_*$ is the tortoise coordinate.
Here, $t_0$, $\theta_0$, and $\phi_0$ are integration constants.
Hereafter, we set these integration constants to zero.

The inhomogeneous solution to the frequency-domain radial Teukolsky equation is given by
\begin{align}
    R(r) = \int_{r_+}^{\infty} \mathrm{d}r'\, G(r,r') T(r'),
\end{align}
where $T(r')$ denotes the source term, and $G(r,r')$ is the Green's function.
For an observer located at infinity, the inhomogeneous solution reduces to
\begin{align}
    R(r\to \infty) = Z r^3 e^{i\omega r_*},
    \label{eq:Ratinfinity}
\end{align}
where 
\begin{align}
    Z(\omega) = \frac{1}{2i\omega B^\mathrm{inc}}\int_{r_+}^\infty \mathrm{d}r' R^\mathrm{in}(r') T(r') \Delta^{-2}(r'),
    \label{eq:defZ}
\end{align}
and $R^\mathrm{in}$ is the ``in'' homogeneous solution satisfying the boundary condition as
\begin{align}
    R^\text{in} &\to
    \begin{cases}
        \Delta^{2} e^{-ikr_*} & \text{for } r\to r_+,\\
        B^\mathrm{ref} r^{3} e^{i\omega r_*} + B^\mathrm{inc} r^{-1} e^{-i\omega r_*} & \text{for } r\to \infty,
    \end{cases}\label{eq:Rinasymp}
\end{align}
with $\Delta = r^2 - 2 M r + a^2$ and $k = \omega - m\Omega_\mathrm{H}$.

We now evaluate $Z(\omega)$ in the near-horizon limit $\epsilon\ll 1$.
Substituting the geodesic trajectories and the asymptotic ``in'' solution~\eqref{eq:Rinasymp} into Eq.~\eqref{eq:defZ}, the leading-order expression reads~\cite{Mino:2008at}
\begin{align}
    Z(\omega) \simeq \tilde{Z}(\omega) \int\mathrm{d}t' e^{i\omega t' - im\phi(t')} e^{-ikr_*(t')} \epsilon(t'),
    \label{eq:Zathorizon}
\end{align}
where $r(t')$ and $\phi(t')$ represent the coordinates of the plunging particle at the time $t'$.
Substituting Eqs.~\eqref{eq:horizongeodesicr} and \eqref{eq:horizongeodesicrstar} into the integral yields the horizon mode pole as
\begin{align}
    \int_{t_\mathrm{i}}^\infty \mathrm{d}t' e^{i\omega t' - im\phi(t')} e^{-ikr_*(t')} \epsilon(t') = \frac{i}{2}\frac{e^{2i(\omega-\omega_1^\MM)t_\mathrm{i}}}{\omega - \omega^\mathrm{MM}_{1}},
    \label{eq:HorizonMode}
\end{align}
where $t_\mathrm{i}$ is the time such that $\epsilon(t_\mathrm{i})\ll 1$. 
Consequently, the frequency-domain inhomogeneous solution $R$ in \eqref{eq:Ratinfinity} possesses the fundamental Matsubara pole, which would give rise to the horizon mode, in addition to QNM poles.

However, Zimmerman and Chen~\cite{Zimmerman:2011dx} demonstrated that the horizon mode is canceled by the Matsubara zero.
They identified an error in the evaluation of $\tilde{Z}$ in Ref.~\cite{Mino:2008at}.
Evaluating $\tilde{Z}$ correctly, we have~\footnote{We follow the conventional definition of $\tilde{Z}(\omega)$ used in Refs.~\cite{Mino:2008at,Zimmerman:2011dx}, which differs from the one used in Ref.~\cite{Oshita:2025qmn}.}
\begin{align}
    \tilde{Z}(\omega) &= \mu \sqrt{\frac{2}{\pi}}\frac{1}{2i\omega B^\mathrm{inc}} \frac{\kappa^3M}{r_+(E-\Omega_\mathrm{H}L_z)} \nonumber \\
    &\quad\times\left[1-3i\frac{kr_+}{\kappa}-2\left(\frac{kr_+}{\kappa}\right)^2\right] \left(aE-\frac{L_z}{\sin\theta_\mathrm{o}}\right)^2\nonumber \\
    &\quad\times\frac{r_+ - ia\cos\theta_\mathrm{o}}{r_+ + ia\cos\theta_\mathrm{o}}\sin^2\theta_\mathrm{o} S_{lm}(\theta_\mathrm{o}),
    \label{eq:Ztilde}
\end{align}
where $E$ and $L_z$ respectively denote the energy and $z$-component of angular momentum per unit mass $\mu$ of the plunging particle.
$\theta_\mathrm{o}$ is the polar angle of the observer. 
The polynomial in the square brackets has zeros precisely at the first two Matsubara frequencies, $\omega^\mathrm{MM}_{1}$ and $\omega^\mathrm{MM}_{2}$, as 
\begin{align}
    & 1-3i\frac{kr_+}{\kappa}-2\left(\frac{kr_+}{\kappa}\right)^2 \nonumber \\
    &= -2\left(\frac{r_+}{\kappa}\right)^2(\omega - \omega_{1}^\mathrm{MM})(\omega - \omega_{2}^\mathrm{MM}).
    \label{eq:sourceprefactor}
\end{align}
This zero at $\omega=\omega^\mathrm{MM}_{1}$ cancels out the pole~\eqref{eq:HorizonMode}, leading to the screening of the fundamental horizon mode.

\begin{figure}[t]
    \centering
    \includegraphics[width=0.9\linewidth]{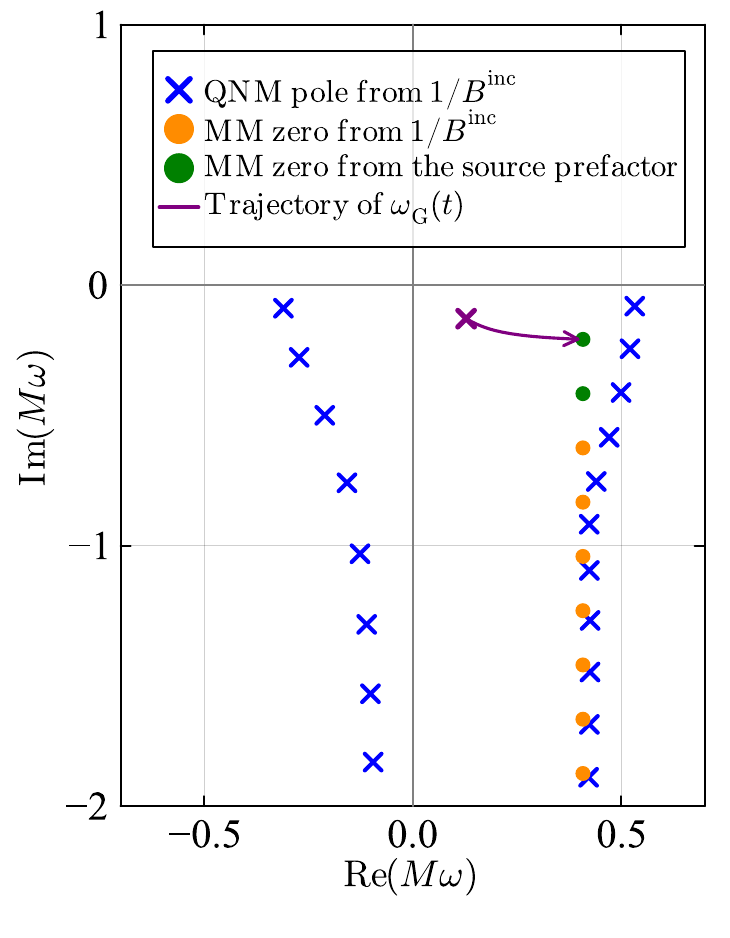}
    \caption{Analytic structure of $\tilde{Z}(\omega)$ for $a/M=0.7$. The blue cross markers denote the QNM poles. The green and orange circles denote Matsubara zeros from the source prefactor~\eqref{eq:sourceprefactor} and the inverse asymptotic amplitude $1/B^\mathrm{inc}$, respectively. The purple solid curve denotes the trajectory of the instantaneous frequency $\omega_\mathrm{G}$ in Eq.~\eqref{eq:instantaneousfreq}.}
    \label{fig:pole_structure}
\end{figure}

\secl{Time-domain waveform and the infinite Matsubara zeros}It is important to note that the source prefactor~\eqref{eq:sourceprefactor} captures only a subset of the zeros of $\tilde{Z}(\omega)$, which we should retain for a complete description.
As illustrated in Fig.~\ref{fig:pole_structure}, the complete analytic structure of $\tilde{Z}(\omega)$ contains an infinite sequence of Matsubara zeros: the first two arise from the source prefactor~\eqref{eq:sourceprefactor}, while those with $j\geq3$ arise from $1/B^\mathrm{inc}$~\cite{Motohashi:2026mbn}.
We can rewrite $\tilde{Z}$ as
\begin{align}
\tilde{Z}(\omega) = \Xi(\omega)\frac{\Pi(\omega)}{Q(\omega)} 
= \Xi(\omega) \frac{\prod_{j=1}^{\infty}(\omega - \omega^\mathrm{MM}_j)}{\prod_{n}(\omega - \omega^\mathrm{QNM}_n)},
\end{align}
where $\Pi(\omega)$ and $Q(\omega)$ possess the Matsubara zeros $\omega^\mathrm{MM}_j$ ($j= 1,2,\ldots$) and QNM zeros $\omega^\mathrm{QNM}_n$ of ordinary ($n= 0,1,\ldots$) and mirror modes ($n= -0,-1,\ldots$), respectively, and $\Xi(\omega)$ has neither poles nor zeros at the Matsubara or QNM frequencies.
As we show below, retaining this full Matsubara-zero sequence leads to the exact cancellation of all source-induced horizon-mode contributions.

The time-domain waveform is obtained by the inverse Fourier transform of Eq.~\eqref{eq:Ratinfinity}. 
To isolate the question of whether source-induced horizon-mode contributions survive, we consider the Matsubara sector of the inverse Fourier transform,
\begin{align}
    h(t) &= \mathcal{F}^{-1} \left[ \Pi(\omega) \int_{t_\mathrm{i}}^\infty \mathrm{d}t' e^{iP(t',\omega)} \right], \notag\\
    &= \int_{-\infty}^\infty \mathrm{d}\omega e^{-i\omega t} \Pi(\omega) \int_{t_\mathrm{i}}^\infty \mathrm{d}t' e^{iP(t',\omega)},
    \label{eq:directwave}
\end{align}
where
\begin{align}
    P(t',\omega) &= \omega t' - m \phi(t') -kr_*(t') +(\kappa/r_+) i t'.
    \label{eq:defP}
\end{align}
The omitted factor $\Xi(\omega)/Q(\omega)$~\footnote{Note that the redshift modes~\cite{DeAmicis:2025xuh,Rosato:2026moe,DeAmicis:2026wqd} originate from this omitted factor. We do not consider the redshift modes nor their cancellation~\cite{Kuntz:2026xep} in this Letter.} is regular and nonzero at the Matsubara frequencies.
It can modify the smooth amplitude and encode the ordinary QNM response, but cannot create or remove a Matsubara pole.
Therefore, it does not affect the cancellation of the source-induced horizon-mode contributions established below~\footnote{An exceptional case occurs when a QNM pole coincides with a Matsubara zero, leading to pole skipping~\cite{Grozdanov:2017ajz,Blake:2017ris,Blake:2018leo,Grozdanov:2018kkt,Kubota:2026hdv}.
In this case, the QNM part contains a contribution corresponding to a Matsubara frequency.}.

\secl{Analysis with saddle-point approximation}In the ``direct wave'' construction proposed in Ref.~\cite{Oshita:2025qmn}, the integral in Eq.~\eqref{eq:directwave} is evaluated using the SPA.
Applying the SPA as in Ref.~\cite{Oshita:2025qmn}, one would obtain
\begin{align}
    h(t) &\simeq \frac{\Pi(\omega_*)}{\sqrt{|H(t'_*)|}} e^{-i\int^{t'_*}\mathrm{d}t' (1+\beta(t'))\omega_\mathrm{G}(t')} .
    \label{eq:directwaveSPA}
\end{align}
Here, $(t'_*, \omega_*)$ is a saddle point at which the saddle-point conditions $\partial_{t'}\Phi=\partial_{\omega}\Phi=0$ are satisfied for the phase $\Phi(t',\omega) =-\omega t + P(t',\omega)$: 
\begin{align}
    t'_* &= t + r_*(t'_*), \\
    \omega_* &=\omega_\mathrm{G}(t') = \frac{1}{1+\beta}\left(m\Omega+m\Omega_\mathrm{H}\beta - i \frac{\kappa}{r_+}\right),
    \label{eq:instantaneousfreq}
\end{align}
where $\beta(t')=-\mathrm{d}r_*/\mathrm{d}t'$ and $\Omega(t')=\mathrm{d}\phi/\mathrm{d}t'$. 
The purple curve in Fig.~\ref{fig:pole_structure} shows the trajectory of $\omega_\mathrm{G}$ for a plunging particle with $L_z=0$ and $E=1$.
$H$ is the determinant of the Hessian matrix defined by
\begin{align}
    H = \left. \det 
    \begin{pmatrix}
    \partial_{t'}\partial_{t'}\Phi & \partial_{t'}\partial_{\omega}\Phi  \\
    \partial_{\omega}\partial_{t'}\Phi  & \partial_{\omega}\partial_{\omega}\Phi 
    \end{pmatrix}
    \right|_{(t',\omega)=(t'_*,\omega_*)}
    = - (1 + \beta(t'_*))^2.
    \label{eq:Hessian}
\end{align}

This SPA implies that Eq.~\eqref{eq:directwaveSPA} reduces to the Matsubara mode with $j=2$ in the horizon limit.
In this limit, the factor $e^{-i\int^{t'_*}\mathrm{d}t' (1+\beta(t'))\omega_\mathrm{G}(t')}$ reduces to the Matsubara mode with $j=1$ as
\begin{align}
    e^{-i\int^{t’_*}\mathrm{d}t' (1+\beta(t'))\omega_\mathrm{G}(t')} \to e^{-i\omega^\MM_1 t},
    \label{eq:phasefactor}
\end{align}
where we use $\beta \to 1$, $\omega_\mathrm{G}(t') \to \omega^\MM_1$, and $t \simeq 2t'_*$.
Since $\omega_\mathrm{G}(t') = \omega^\mathrm{MM}_{1} + \mathcal{O}(\epsilon(t'))$ with $\epsilon(t') = e^{2 \mathrm{Im}(\omega^\mathrm{MM}_{1}) t'}$ and $t \simeq 2t'_*$, the factor $(\omega_\mathrm{G}(t'_*) - \omega^\mathrm{MM}_1)$ contained within $\Pi(\omega_*)$ acts as an exponential damping factor
\begin{align}
    (\omega_\mathrm{G}(t'_*) - \omega^\MM_1) \propto e^{\mathrm{Im}(\omega^\mathrm{MM}_{1}) t}.
    \label{eq:firstMatsubarazero}
\end{align}
In contrast, the other factors in $\Pi(\omega)$ do not exhibit exponential decay, behaving instead as $(\omega_\mathrm{G}(t'_*) - \omega^\MM_j) \simeq -i\mathrm{Im}(\omega^\MM_1)(j-1)$ for $j\geq 2$.
The Hessian becomes $H=-(1+\beta)^2\to -4$.
Consequently, under the SPA, Eq.~\eqref{eq:directwaveSPA} would yield~\cite{Oshita:2025qmn}
\begin{align}
    h(t) \propto e^{-i\omega^\MM_2 t}.
    \label{eq:resultSPA}
\end{align}
The SPA result thus suggests that the corresponding contribution would be masked by the longer-lived QNM response, motivating the expectation that QNM filtering could expose it as a residual attributed to near-horizon emission.

In the literature, only one paper~\cite{Kuntz:2026xep} has explicitly questioned the SPA-based construction in the Schwarzschild case~\footnote{Relatedly, Ma and Wang characterized the SPA-based derivation as heuristic, since its complex-frequency contour deformation was not specified~\cite{Ma:2026hcb}.
Nevertheless, their alternative treatment of the filtered waveform obtains a contribution governed by the same $\omega_\mathrm{G}$ from the ``anti-causal'' poles introduced by the filter, which they interpret as a ``direct wave'' pattern.
This interpretation contrasts with the filter-redistribution analysis of Ref.~\cite{Cheung:2026gfd}, which finds that the corresponding filtered features are dominated by radiation sourced at or outside the light ring.}.
Kuntz and Della Rocca noted that the saddle point lies at the endpoint of the causal source integral, so that the standard interior-saddle treatment is inapplicable~\cite{Kuntz:2026xep}.
Still, no study has evaluated the Kerr source integral underlying Eq.~\eqref{eq:directwave} without invoking the SPA.
Here, Eqs.~\eqref{eq:phasefactor} and \eqref{eq:firstMatsubarazero} explicitly show that the factor associated with the first Matsubara zero in the prefactor $\Pi(\omega_*)$ varies on the same time scale as the exponential phase factor.
The assumed separation between a slowly varying prefactor and a rapidly varying phase in the SPA is therefore not controlled in the late-time near-horizon limit considered here.

\secl{Exact cancellation from the infinite tower of Matsubara zeros}Now we evaluate the near-horizon source integral~\eqref{eq:directwave} without invoking the SPA, and establish that it exhibits the exact cancellation of all source-induced horizon-mode contributions.
Using the geodesic equations near the horizon~\eqref{eq:horizongeodesicr}--\eqref{eq:horizongeodesicrstar}, the phase $P$ in \eqref{eq:defP} can be expressed as
\begin{align}
    P(t',\omega) = 2kt' +(\kappa/r_+) i t' + \sum_{n=1}^\infty b_n \epsilon^n(t'). \label{eq:phaseP}
\end{align}
Here, we omit the constant factor.
The specific forms of the expansion coefficients $b_n$ are not required for this analysis. 
Then the integrand of Eq.~\eqref{eq:directwave} yields
\begin{align}
    e^{iP} = e^{ i (2 k +(\kappa/r_+) i  )t'}\sum_{n=0}^\infty \tilde{b}_n \epsilon^n(t'),
\end{align}
with $\tilde{b}_n$ being redefined coefficients. 

Recalling Eq.~\eqref{eq:horizongeodesicr}, integrating the $n$-th order term yields the $(n+1)$-th Matsubara pole as
\begin{align}
    & \int \mathrm{d}t' e^{ i (2 k +(\kappa/r_+) i  )t'} \epsilon^n(t') 
    \nonumber \\
    & = \int \mathrm{d}t' e^{i (2 k +(n+1)(\kappa/r_+) i )t' } \propto \frac{1}{\omega - \omega^\MM_{n+1}}.
\end{align}
Consequently, the integral yields the sum of the Matsubara poles
\begin{align}
    \int \mathrm{d}t' e^{iP} = \sum_{n=0}^\infty \bar{b}_n \frac{1}{\omega - \omega^\MM_{n+1}}, \label{eq:MMpolesum}
\end{align}
where $\bar{b}_n$ are also redefined coefficients. 
These poles are canceled by the infinite tower of Matsubara zeros in $\Pi(\omega)$.

In the SPA-based result~\eqref{eq:directwaveSPA}, the frequency-dependent factor $\Pi(\omega)$ is replaced by its saddle-point value $\Pi(\omega_*)$, so the sequential pole--zero cancellations are not retained.
By contrast, retaining $\Pi(\omega)$ in the exact frequency integral cancels every source-induced Matsubara pole.
Therefore, the exact analysis without SPA shows that all source-induced horizon modes vanish.

Although the coefficients $b_n$ in Eq.~\eqref{eq:phaseP} depend on the Kerr parameters and on the orbital data, they affect only the coefficients $\bar{b}_n$ in Eq.~\eqref{eq:MMpolesum}.
For any plunging trajectory for which the near-horizon expansion applies, each would-be source-induced Matsubara pole is canceled by the corresponding zero in $\Pi(\omega)$.
The cancellation mechanism is therefore independent of the detailed orbital profile and of the values of the black-hole and trajectory parameters.

\begin{figure}[t]
    \centering
    \includegraphics[width=0.9\linewidth]{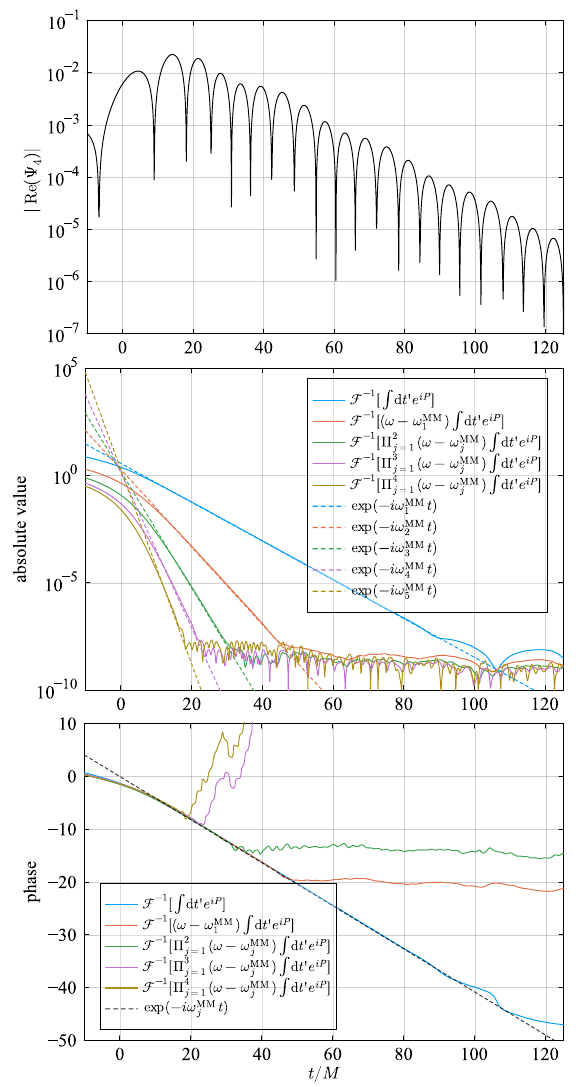}
    \caption{Time-domain evolution of source-induced horizon-mode contributions for a particle with $E=1$ and $L_z=0$ plunging into a Kerr black hole with $a/M=0.7$: waveform (top), absolute value (middle), and phase (bottom).
    We choose $t = 0$ to correspond to $\epsilon = 1$. 
    The solid curves show the unscreened source contribution (blue) and the contributions including Matsubara zeros through $\omega^\MM_1$ (orange), $\omega^\MM_2$ (green), $\omega^\MM_3$ (purple), and $\omega^\MM_4$ (brown).
    The dashed curves are damped sinusoids at the corresponding Matsubara frequencies. 
    Each additional zero suppresses the candidate contribution at the next Matsubara frequency. 
    The SPA stops after the first zero and hence predicts a residual contribution at $\omega^\MM_2$ (orange), whereas retaining the full infinite tower yields the exact cancellation established in the text.}
    \label{fig:waveform_suppression}
\end{figure}

\secl{Numerical verification}To confirm the validity of this analytical mechanism, we numerically demonstrate that the first four horizon modes vanish. 
To regulate the numerical inverse Fourier transform~\eqref{eq:directwave}, we multiply the integrand by the smooth window function
$w(M\omega)= \{1 - \tanh[((M\omega)^2- 4^2)/4]\}/2$, which suppresses high-frequency contributions~\footnote{We verified that the qualitative behavior is unchanged under variations of the window parameters and upon using alternative smooth windows, including the Tukey and Planck-taper windows.}.

Figure~\ref{fig:waveform_suppression} illustrates the sequential screening of the time-domain horizon-mode contributions for a plunging particle.
The unscreened source contribution (blue solid) approaches the fundamental Matsubara frequency $\omega^\mathrm{MM}_1$ (blue dashed).
Including the corresponding zero removes this candidate contribution and yields the orange curve, which approaches $\omega^\mathrm{MM}_2$.
The SPA treatment retains only this first screening stage and therefore predicts a residual contribution at $\omega^\mathrm{MM}_2$.
The exact calculation instead retains the second and higher Matsubara zeros: the green curve, which includes the first two zeros, is screened at $\omega^\mathrm{MM}_3$, while the purple and brown curves show the subsequent screening at $\omega^\mathrm{MM}_4$ and $\omega^\mathrm{MM}_5$, respectively.
Continuing this sequence through the infinite Matsubara-zero tower yields the exact cancellation of all source-induced horizon-mode contributions.

Since the near-horizon analysis applies when $\epsilon< 1$, i.e., $t/M> 0$, at least in this setup, the exact cancellation of horizon modes applies in this regime even before the waveform peak at $t/M\simeq 14.1$ and the prograde light ring crossing at $t/M\simeq 6.2$, corresponding to the QNM response.

\secl{Conclusion}We have reexamined the late-time horizon-mode limit assumed in the proposed ``direct wave'' construction for a point particle plunging into a Kerr black hole.
That construction invokes the saddle-point approximation (SPA) to evaluate the source integral, but the approximation is not controlled in the near-horizon limit.
An exact evaluation of the source integral instead reveals that an infinite tower of Matsubara zeros sequentially cancels all the would-be horizon-mode contributions.
Consequently, all source-induced horizon-mode contributions vanish within this near-horizon perturbative calculation.

Our result does not, by itself, determine the origin of the oscillatory features in QNM-filtered plunging-particle or comparable-mass numerical-relativity waveforms.
It does establish that a plunge-sourced contribution cannot asymptote to a surviving horizon mode in the near-horizon perturbative problem considered here.
It also reveals the hidden analytic mechanism by which an infinite tower of Matsubara zeros in the near-horizon source response eliminates every would-be late-time horizon mode.
The cancellation mechanism holds independently of the detailed orbital profile and of the values of the black-hole and trajectory parameters.
The proposed ``direct wave'' interpretation based on that late-time horizon-mode limit is therefore ruled out.

\vspace{3mm}
\secl{Acknowledgments}
We thank Emanuele Berti, Mark Ho-Yeuk Cheung, Paolo Pani, Romeo Felice Rosato, and Sophia Yi for useful discussion.
This work was supported in part by JSPS KAKENHI Grant No.~JP22K03639 and No.~JP26K21813 (H.M.). This research was supported in part by Perimeter Institute for Theoretical Physics. Research at Perimeter Institute is supported in part by the Government of Canada through the Department of Innovation, Science and Economic Development and by the Province of Ontario through the Ministry of Colleges and Universities.

\bibliography{ref}
\end{document}